# Thermal Rectification from Size-Dependent Phonon Confinement in Nanoparticle Assemblies

Megan J. Farrington[1 2], David J. Wasserman[1 2], Eden R. Josza[1 2], Alessandro Podestà[1], Alessio Zaccone[1], Mariia Sidorova[3 4], Alexej D. Semenov[3], Nicola Ludwig[1] and Marcel Di Vece[1 2*]

[1] *Physics Department "Aldo Pontremoli", Università degli Studi di Milano, Via Celoria 16, 20133, Milan, Italy*

[2] *Interdisciplinary Centre for Nanostructured Materials and Interfaces (CIMaINa), Via Celoria 16, 20133, Milan, Italy*

[3] *German Aerospace Centre (DLR), Institute of Space Research, Rutherford's. 2, 12489 Berlin, Germany*

[4] *Physikalisch-Technische Bundesanstalt (PTB) Berlin, Abbestraße 2–12, 10587 Berlin, Germany*

[*] *Corresponding author: marcel.divece@unimi.it*

## Abstract

Thermal insulation remains an important technological challenge across the vast number of applications, from living quarters to quantum technology. Here, we exploit the size-dependent modification of the phonon density of states arising from phonon confinement in nanoparticles to fabricate a simple phonon rectifier. The smaller of the two connected nanoparticles imposes stronger phonon confinement leading to rectifying phonon transport. This concept is extended to the macroscale by constructing two overlapping layers of differently sized nanoparticles, thereby realizing a macroscopic phonon diode. Following the localized heat deposition by laser light, the temperature profiles across a phonon-diode were measured by infrared imaging. Although the rectifying strength is moderate, the abundance of optimization possibilities makes this method promising for ultra-low volume thermal insulation at both the nano- and macroscale.

The need to control heat transport extends from advanced electronic devices such as microprocessors to providing agreeable temperatures in living quarters with a technologically more sensational application: thermal camera cloaking devices [1,2]. Thermal management, which would ideally be achieved by passive methods with the advantage of negligible energy consumption, require the ability to block heat transfer, analogous to the electronic counterpart: the diode [3]. Thermal rectification can be suitably achieved by controlling radiation [4] or phonon diffusion using insulators or semiconductors since they do not have free electrons to carry heat and therefore rely on lattice vibrations only [5]. The phonon rectifier, i.e. a thermal diode [6,7], could suppress the flow of heat in one direction, enabling the passive stabilization of the temperature of an object [8]. This can for example be realized by nonlinear lattices [9,10] and anharmonic graded mass crystals [11], that could lead to complicated thermal devices [12]. Extending this to nanophononics, thermal management at the nanoscale, which can for example also be achieved by phonon crystals [13], can lead to intricate ways to manipulate heat such as for example a phonon laser [14].

In analogy to quantum confinement in semiconductors [15–19], geometric confinement opens a gap in the occupied phonon momenta of nanoparticles [20]. Confinement reduces the number of available degrees of freedom, thereby modifying the phonon density of states (PDOS) [21] and leading to a crossover from the Debye $\omega^2$ scaling to an anomalous $\omega^3$. This modified PDOS, together with phonon boundary scattering, leads to reduced thermal transport [22], as was found to depend on the diameter of silicon nanowires with two orders of magnitude reduced thermal conductivity [23,24]. Similar effects have been predicted and observed for thin films [21]. By connecting two nanoparticles of different sizes, a certain region of the PDOS in each nanoparticle becomes effectively gapped, i.e. inaccessible for phonons traveling from the other nanoparticle [25]. Phonon wavelengths larger than the gap cannot be excited in the heat-receiving particle, which reduces the heat transfer. The reduction is different for opposite directions of the heat flow, in effect creating a phonon diode [22,26]. This removes the necessity for using bulky materials with low thermal conductivity as thermal insulators, instead a particle size-engineered gapped PDOS will suffice. In nanostructures, the main phonon transport mechanism is transitioning from diffusive to ballistic [27,28], in which long-wavelength phonons remain ballistic even with grain boundaries [29]. This creates a local temperature differences, which could enhance the thermal rectification.

In general, the crystal structure allows phonon wavelengths range from ~1 nm at the Brillouin zone edge to the μm scale near the Γ-point ($k \approx 0$) s. However, due to energy considerations, at room temperature the thermal phonons have wavelengths extending to ~10 nm [30,31] which are comparable to the size of the nanoparticle, thus making germanium an excellent choice due to its moderate phonon dispersion [32,33]. Because the Debye frequency is proportional to $T_D$, the phonon spectrum in germanium indeed spans a smaller frequency interval across the Brillouin zone because it has a much lower Debye temperature ($T_D$) than Si ($T_D$≈400 K vs ≈700 K). For the same particle size, the confinement-induced low-k cutoff excludes only a slightly larger fraction from the Debye sphere in germanium than in silicon, meaning that the expected confinement in germanium is only slightly stronger than in silicon.

The effect of confinement disables a fraction of the PDOS, starting at the lowest k-values and up to a k value which depends on the nanoparticle size. With the germanium PDOS in a narrow k-space range, at relatively lower phonon frequencies as compared to e.g. silicon [34], the confinement quickly covers a substantial part of the PDOS k-range, making it an excellent material for size dependent phonon confinement. Restricted heat transport at those phonon wavelengths could potentially provide close to 100% thermal rectification at room temperature [26].

The PDOS of a germanium nanoparticle was experimentally found to significantly depend on the size in the range <10 nm: by decreasing the particle size, a decreased PDOS in the low frequency range was found and vice versa [32], confirming the potential to develop phonon rectifying properties; a phonon with a certain frequency can only be transmitted to the nearby particle if there exists a corresponding vibrational state on the opposite side with the same frequency [35].

In this work, the phonon diode is realized by the connection between a small and a larger germanium nanoparticle (Fig. 1a left), creating a PDOS mismatch as shown in k-space in Fig. 1a (right) where the Debye sphere with an empty spherical slab results in a direction-dependent restriction of the phonon transport between nanoparticles as also explained in Semenov et al. [26]. The connection between the small and larger nanoparticle is extended over a larger, macroscale overlap region of two nanoparticle assembled layers, each consisting of particles of different size (Fig. 1b). This novel method, echoing the role of germanium in the first transistor [36], has the potential for the fabrication of highly insulating and low volume and low cost materials with considerable energy saving results.

The germanium nanoparticles without binder molecules [35,36], leading to a well-defined PDOS and connectedness, routinely fabricated in previous studies [37–39], using a gas aggregation plasma sputtering nanoparticle generator [40,41], form a cauliflower structure [37,38]. This structure was corroborated by atomic force microscopy (AFM) of the germanium nanoparticle assembled films (Fig. 2a-b and Fig. S5) and was confirmed by scanning transmission electron microscopy (STEM) (SI Fig. S4a-b). They exhibit small spherical features of ~7nm within larger ~50-70 nm structures. This cauliflower structure implies that the porosity of the nanoparticle assembled film consists of the space between and within the cauliflower-like nanoparticles, which cannot be described by a well-defined random close packing model [42], providing ample contact area for heat transport.

Two different germanium nanoparticle sizes were achieved by using an aggregation distance of 30 and 70 cm, resulting in small particles, leading to respective films with rms roughness Rq = 53.2 ± 1.3 nm and in larger particles films with Rq = 71.6 ± 2.4 nm. Although electron diffraction has shown the presence of predominantly crystalline germanium, a small amorphous fraction was observed [43], which was confirmed by Raman spectroscopy (near the Γ-point [44]) (SI Fig. 5). The amorphous interlayer facilitates interfacial thermal transport due to improved phonon density of states overlap and more delocalized interfacial modes [45].

The phonon rectifying properties were characterized by thermal imaging (Fig. 1c) using the method described by Golovin et al. [46], resembling thermal modulated thermo-reflectance [47], resulting in heat profiles of the germanium nanoparticle assembled film induced by a laser point heat source. The thermal image clearly exhibits a lateral heat flow as indicated by the radial temperature gradient shown in Fig. 1c. The $\Delta T(x)$ profiles along the midline of the flattening window (SI Experimental section), originating at the laser spot, are shown in Fig. 3a and b, for a uniform germanium and gold nanoparticle assembled layer, respectively. The temperature profiles were acquired before (heating phase), during and after (cooling phase) the steady state regime. Because thermal conductivity of the gold (largely electron-mediated) is orders of magnitude larger, the temperature gradient is much smaller and the $\Delta T(x)$ curves are much flatter than those of the germanium nanoparticles layers. The limited connectedness of the nanoparticles at the 10-50 nm scale in these films will reduce the phonon transport length scale of germanium, which is ~200 nm at 300K [48], leading to a smaller thermal conductivity. When the heat remains local, it gives rise to a larger thermal gradient across the sample, while for the large thermal conductivity the temperature profile smooths out quickly.

Taking into account that the infra-red light emission from the glass substrate in the 8–14 μm range originates only from a near surface layer of thickness ~1–10 μm [49], while emission from germanium originates from the full layer thickness [50], the measured temperature is dominated by the glass contribution (85-90%), where only 15-10% of the signal can be attributed to germanium nanoparticles. Since the effective thermal conductivity is expresses as:

1) $$k_{eff} \approx \frac{k_{Ge}t_{Ge} + k_{glass}t_{glass}}{t_{Ge} + t_{glass}}$$

with k the thermal conductivity and t the thickness as measured by the thermal camera, the measured thermal conductivities of the germanium and gold nanoparticle layers are underestimated by a factor of ~12.

To calculate the thermal conductivity of the nanoparticle assembled films for the cooling phase, several heat maps at different times were used, providing independent spatial and temporal information. The 2D thermal diffusion was approximated by the solution of the Fourier equation, yielding averaged thermal conductivities obtained by fitting the experimental profiles with an exponential decay function (proportionally to the square of variable, Eq. S26) for distances larger than the diffusion length [56].

For the germanium nanoparticle assembled films, the extracted thermal conductivities were, 1.8±0.4 W/(m·K) and 2.2±0.1 W/(m·K) for the cooling phase, 2.0±0.1 W/(m·K) and 1.5±0.1 W/(m·K) for the steady state for the larger and smaller nanoparticles, respectively (SI Table S1). Although these values are consistent, during the heating stage, the thermal conductivity of 0.84±0.02 W/(m·K) and 2.3±1 W/(m·K), for the larger and smaller nanoparticles respectively, required a correction factor of 12 to account for the glass substrate contribution. This correction is not required for the cooling phase since the relatively large glass volume ensures a constant background temperature for the changing temperature in the germanium nanoparticle alone. One reason for the difference between the thermal conductivities of the large and small germanium nanoparticles is likely the difference in inter-particle contact surface area, which is relatively larger for the smaller nanoparticles. Another explanation is the reduced integrated PDOS of the smaller nanoparticles due to the expected nanoscale induced confinement [51] as compared with that of the larger nanoparticles.

Since the bulk germanium thermal conductivity is 60 W/(m·K) [52,53], this points to a ~30 times reduced thermal conductivity, likely due to changes in the PDOS, contact area reduction [54,55] and increased presence of defects. This is in agreement with the reported almost two orders of magnitude reduced thermal conductivity of 4nm semiconductor nanoparticles compared with bulk [56].

To determine the thermal conductivity of the glass substrate independently, a small black ink spot was used to deposit the laser energy into the glass, yielding a thermal conductivity of 1.5 ± 0.4 W/(m·K) in agreement with the literature [57]. Although this is considerably lower than that of bulk germanium, its effect on the measurement as a parallel heat conduction path should be considered for the interpretation of the above results because of the much lower effective germanium nanoparticle thermal conductivity. Considering the glass support size, shape, and the specific heat (0.84 J/g·K), together with a heat source of ~60mW, a temperature increase of ~15°C would require ~30 s, assuming radiative and convective heat losses of ~1mW at ΔT= 10° C with the ambient. Since this agrees with experimentally determined values, the glass substrate can be concluded to create an elevated background temperature. Heat from the covering germanium nanoparticle layer may dissipate into the glass mainly by radiation as the thermal resistance is high due to the strongly limited contact surface area [58], enhanced by the cauliflower structure, and likely van der Waals bonding, as observed previously in 2D layered materials [59,60]. As a reference, the thermal conductivity measurements of gold nanoparticles on the same glass substrates yielded 38±4 W/(m·K), an order of magnitude less than the bulk value of 317 W/(m·K)[64], due to the reduced contact area and glass support. The distinction between the gold and germanium nanoparticles confirms the validity of the method used because of the observed proportionality.

As shown in Fig. 3c, the ΔT(x) profiles show clear differences between the germanium, gold, and glass alone with enhanced visibility using the logarithmic ΔT scale. These profiles are well described by the isotropic steady state solution of the Poisson equation in polar coordinates with combined Neumann (at the laser spot) and Dirichlet (at the sample edge) boundary conditions [26] (presented in SI). Due to the lower glass thermal conductivity, it is possible to follow the characteristic downward curving of ΔT (~3.5mm), in agreement with the natural logarithm of the inverse distance (S25 and S30). This downward turn was not observed for the profiles for germanium and gold because their larger thermal conductivities required a longer distance.

The phonon diode is constructed by the overlap of two germanium nanoparticles layers consisting of the two different nanoparticle sizes, as shown in Fig. 1b and Fig. 4a. Temperature profiles are extracted along a line of a pencil mask, crossing the nanoparticle overlap area, in forward and reverse direction of the phonon diode as schematically depicted in Fig. 4b. The clear difference between the left facing (reverse) and right facing (forward) heat flow directions (schematically indicated in Fig. 5a), within the phonon diode are shown in Fig. 5b. For the left facing (reverse) heat flow in the diode, a significant deviation in the temperature profile is observed compared to a uniform layer with ΔT≈0.04° C lower at the entrance of the diode region, extending towards the end of the overlap area. In contrast, for the right facing (forward) heat flow configuration, the temperature is higher with ΔT≈0.04° C and the profile starts to decay slower close to the phonon diode area, as evidenced by the deviation from the non-diode reference (uniform sample) curve.

The directional dependence of the heat flow is explained by the different phonon blocking effects in the heat receiving nanoparticle (Fig. 1a). Indeed, the germanium nanoparticles at the diode left-facing area are larger than those at the right-facing area. The rectifying strength R is defined as:

$$2) \qquad R = \frac{J_{forward} - J_{reverse}}{J_{reverse}} = \frac{\Delta T_{forward} - \Delta T_{reverse}}{\Delta T_{reverse}} \qquad \text{assuming: } J \propto \Delta T$$

with J the heat flux and T the temperature. For R=0 no rectification occurs while large R corresponds to strong rectification. The rectifying strength of ~6% is estimated by comparison with the right-facing (double layer) and un-interrupted phonon flow (i.e. reverse and forward direction) at the end of the phonon diode region, by taking their respective temperature differences. Considering the correction on the thermal infra-red measurement by the effect of glass, the actual ΔT of the germanium nanoparticles is underestimated by about seven times, making it likely that the rectifying strength approaches ~20%. Using the thermal conductivity of the germanium nanoparticle layers, the heat flow through the entire surface of the diode is ~3mW, resulting in ~200 μW blocked heat.

The gradual temperature deviation from a uniform layer indicates that the thermal rectifying occurs equally over the entire extend of the overlap range. While most of the heat in the overlap region is flowing laterally, at each consecutive nanoparticle contact a small portion of the heat is transferred vertically in the large to small particle direction or vice versa. Considering the phonon diode active area, which is ~0.5 $mm^2$ and the individual nanoparticle contact area of a few $nm^2$, the rectifying strength of each nanoparticle contact is likely much larger. The qualitative agreement with the theory of Semenov et al. [26] suggests the validity of the explanation by the phonon-confinement model and that experimental imperfections do not fully hide the phenomenon.

The phonon rectifier was mimicked by solving the Fourier's law in reverse a forward direction by halving and doubling the thickness of the germanium layer, respectively (Fig. 5c). Remarkably, to approach the experimental values, the interface heat transport required $G < 10$ $W/m^2K$, which is much lower than full-contact interfaces [62] and is caused by the considerably reduced nanoparticle-glass contact surface area [58]. In agreement with the experiments, a decrease and increase in ΔT was observed for the phonon diode reverse and forward direction by halving and doubling the thickness of the germanium layer, respectively.

The experimental temperature for the left facing (larger NPs) in Fig. 5b is somewhat lower around the entry of the diode, as compared to the non-diode case (uniform layer). This temperature drop is most likely caused by the local doubling of the nanoparticle layer (=diode), enabling more heat flow as confirmed by the calculations in Fig. 5c. The full regression of ΔT when the layer is back to its original thickness indicates a low amount of heat loss.

In dielectric solids, the thermal conductivity at the nanoscale is often characterized by their doping, defects, grain boundaries, anisotropy and periodicity [63,64]. The effects of imperfections in the germanium nanoparticle assembled layer, such as the presence of (surface) defects [63,64], amorphous portions, size polydispersity, film thickness etc are treated in the SI and considered to remain lower than the phonon diode measurement significance.

The simplicity of this phonon-diode design, consisting only of two nanoparticles of different size of the same material, makes it a promising platform for technological development. By increasing the particle size contrast, the PDOS and phonon transport between nanoparticles can be considerably modified, leading to potentially stronger rectification. Further optimization could proceed by improving crystallinity, engineering the nanoparticle contact area etc. leading to commercial applications of nanoparticle-based phonon diodes for thermal management and thermal camouflaging.

**Acknowledgement**

The authors thank P. Piseri and A. Giugni for Raman spectroscopy assistance, STEM imaging of the nanoparticles by A. Falqui and A. Casu, L. Dal Fabbro, and M. Meneghini for assistance in AFM imaging and S. McCary and L. Kroon for assistance with thermal imaging experiments.

**Author contributions**

MDV conceived the work and had overall supervision. MDV, NL, MJF, DJW performed the thermal experiments and data analysis. The theoretical phonon confinement modelling and analysis was performed by MS, AS, and AZ and thermal diffusion modelling by ERJ, all supporting experiments. AP performed AFM and particle size analysis. MDV wrote the manuscript with discussions and editing from all authors.

**Conflict of Interest**

MJF, DJW, ERJ, AZ, MS, ADS, NL, and MDV filed a patent based on this work.

# Figures

FIG. 1

A)

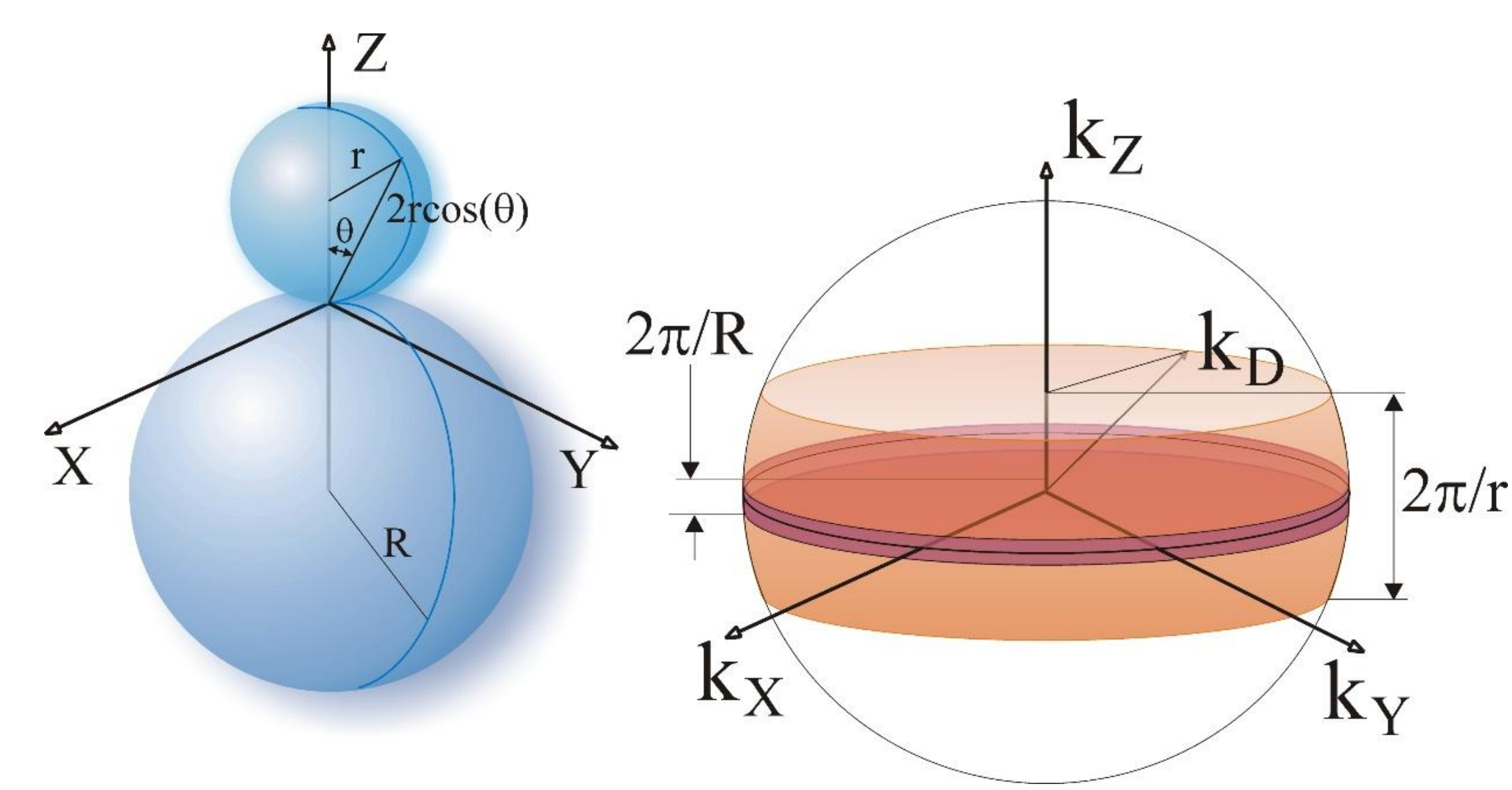

A) Different confinement in spherical nanoparticles with different size. Left side: Real-space schematic of two spherical particles in thermal contact, where the smaller and larger nanoparticle have with radii $r < R$. Right side: confinement in wavevector k-space. The allowed states in the Debye sphere of radius $k_D$ are those populating the sphere outside a spherical slab of thickness $2\pi/r$ (or $2\pi/R$) centred in the origin. That is, the phonons leaving the bigger nanoparticle to propagate into the smaller one can only occupy the states in momentum space outside the slab of thickness $2\pi/r$. This is a much smaller space than for the phonons travelling from the smaller to larger nanoparticle, which can occupy the Debye sphere outside the slab of thickness $2\pi/R$: hence the rectifying properties (see also Semenov et al. ref 26).

B)

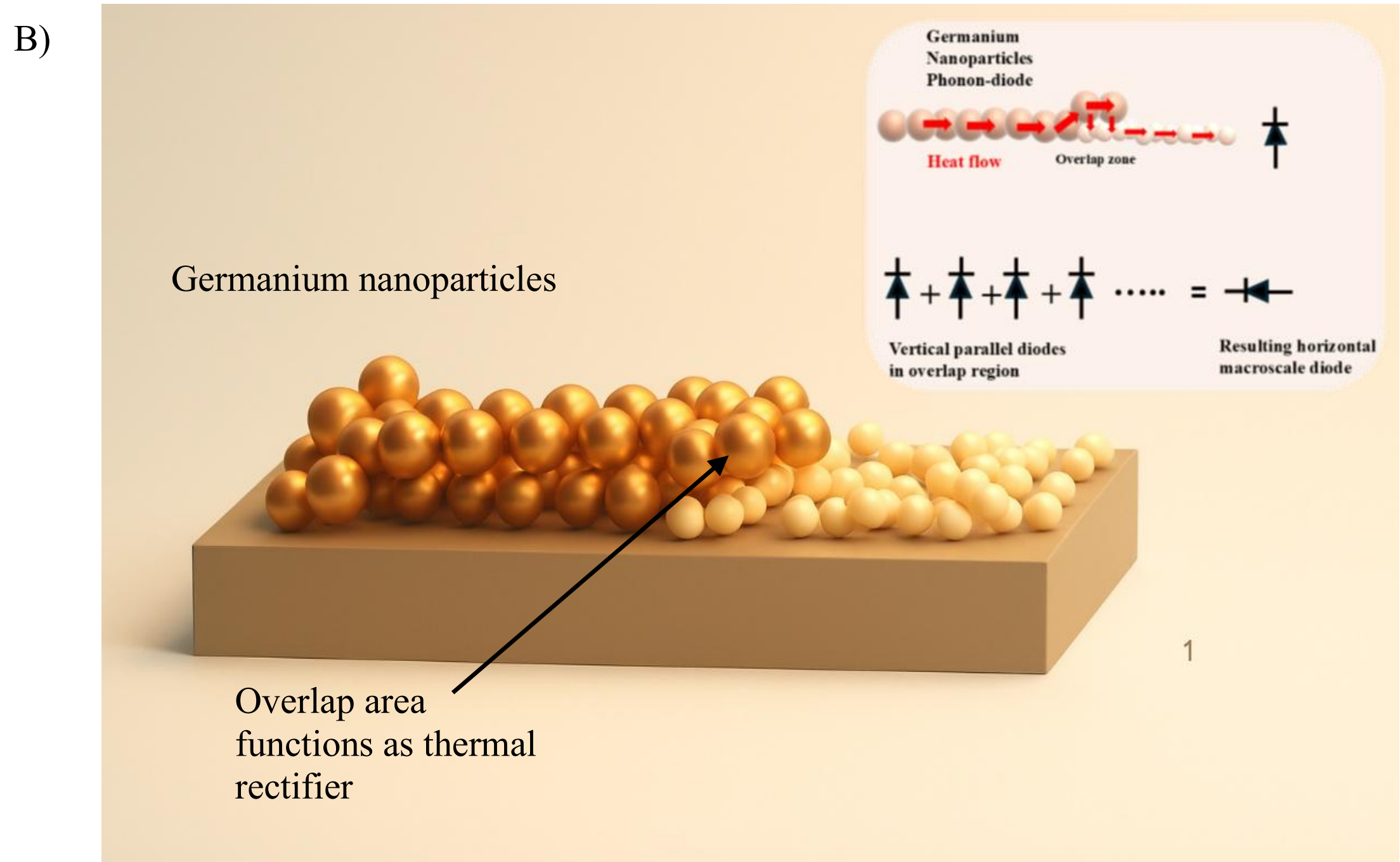

B) A cartoon of the experimental nanoparticle based thermal rectifier concept using nanoparticles of different size. Due to the restricted flux of phonons from the larger into the smaller particle, caused by the mechanism illustrated in panel A, the heat transmitted from large to small is lower compared to the heat transmitted from small to large. By connecting the unequally sized nanoparticles, phonons of a certain energy are therefore blocked in one direction, creating a macroscale (partial) phonon diode emergent from the nanoscale. Inset: schematic depiction of heat flow, vertical parallel diodes between large and smaller nanoparticles and emergent macroscale horizontal diode.

C)

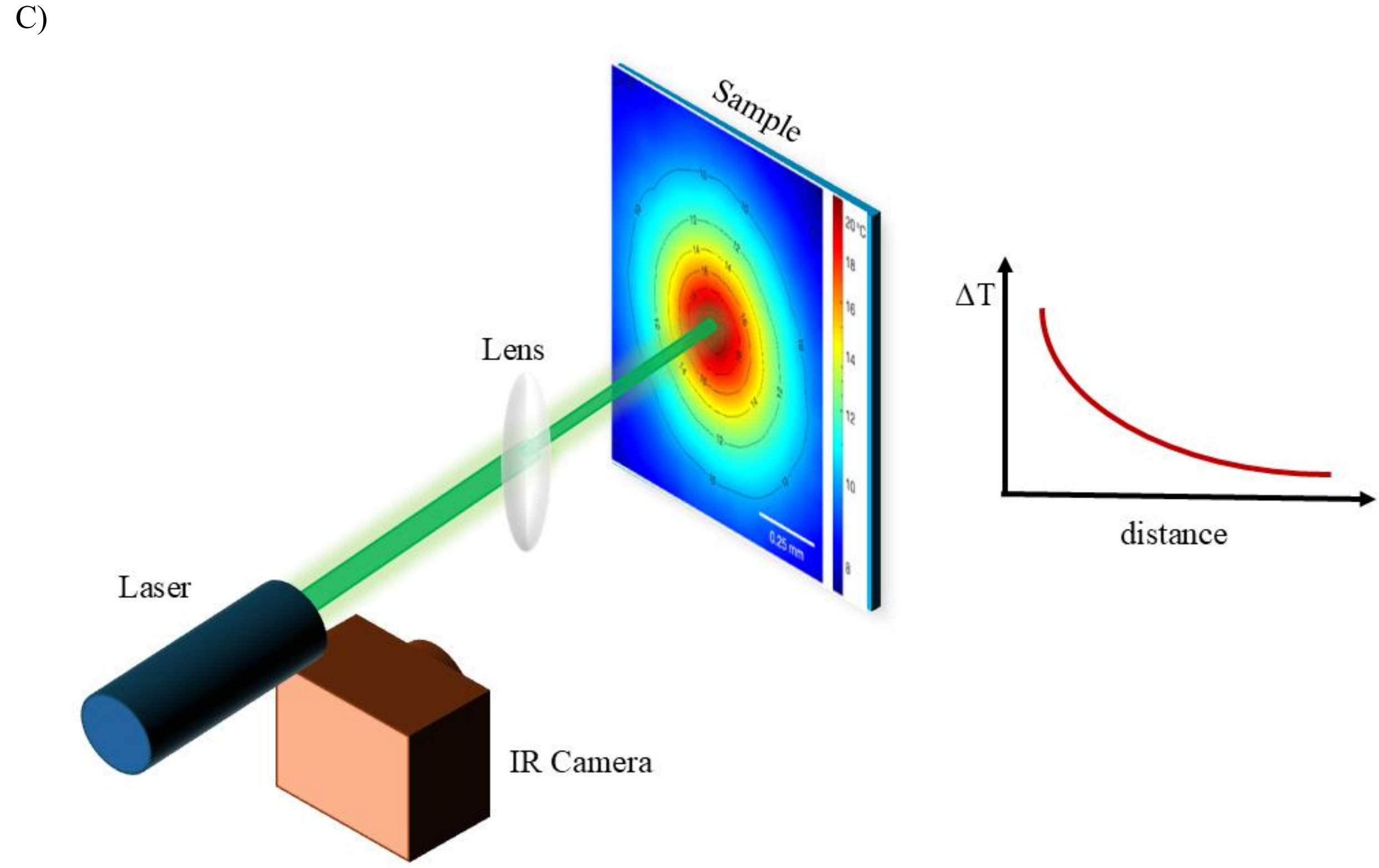

Schematic impression of experimental setup using a green laser with focusing lens to deposit localised heat on the germanium nanoparticle layer sample. The thermal camera monitors the temperature which can increase by up to ~15°C. The thermal recordings as shown in the picture (real data) are analysed to obtain thermal gradient profiles (right graph) across the nanoparticle (overlap) area to determine the thermal conductivity and phonon rectifying characteristics. The colours on the sample indicate the thermal camera recording with each colour corresponding to a temperature.

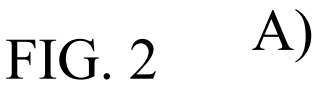

A)

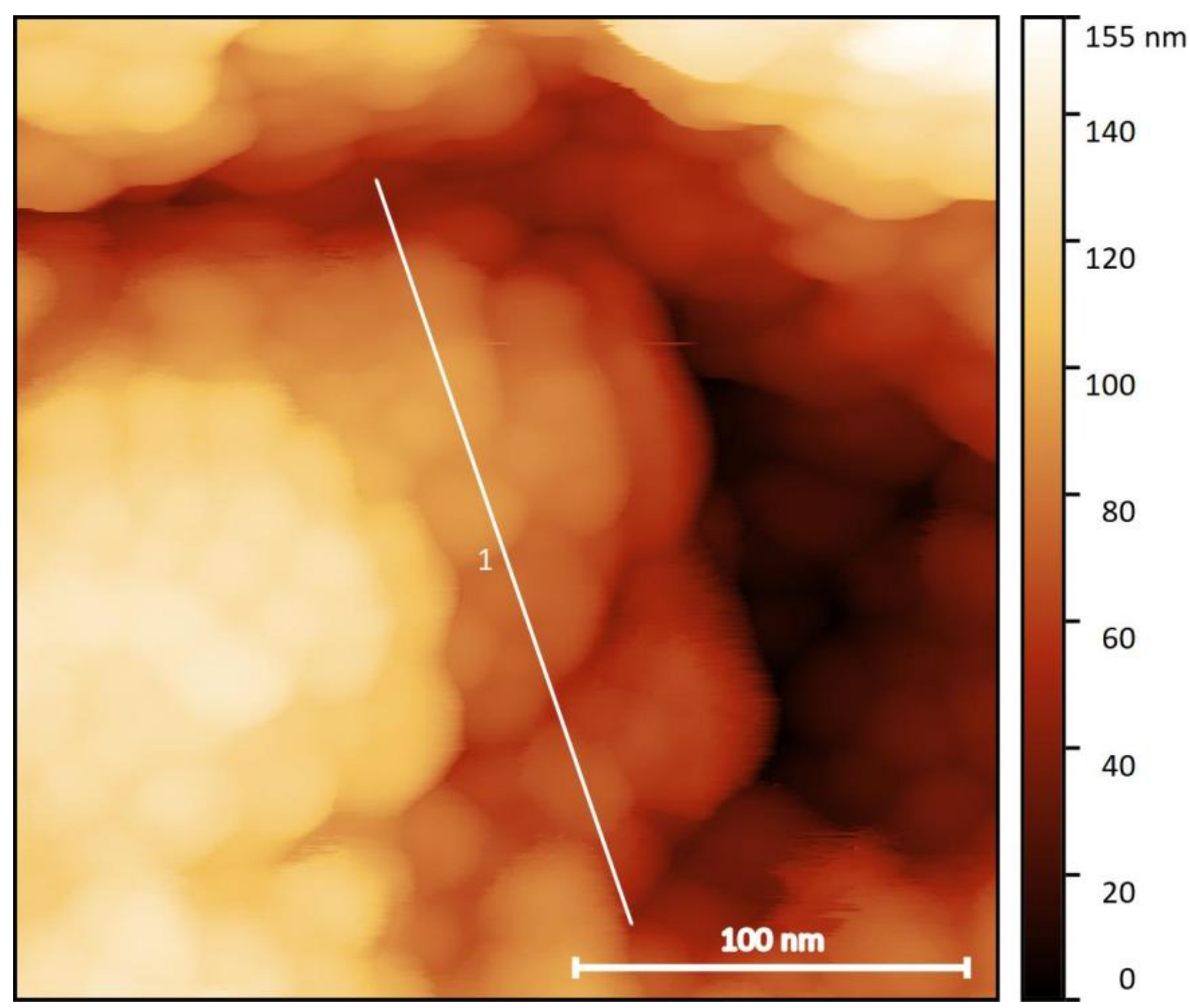

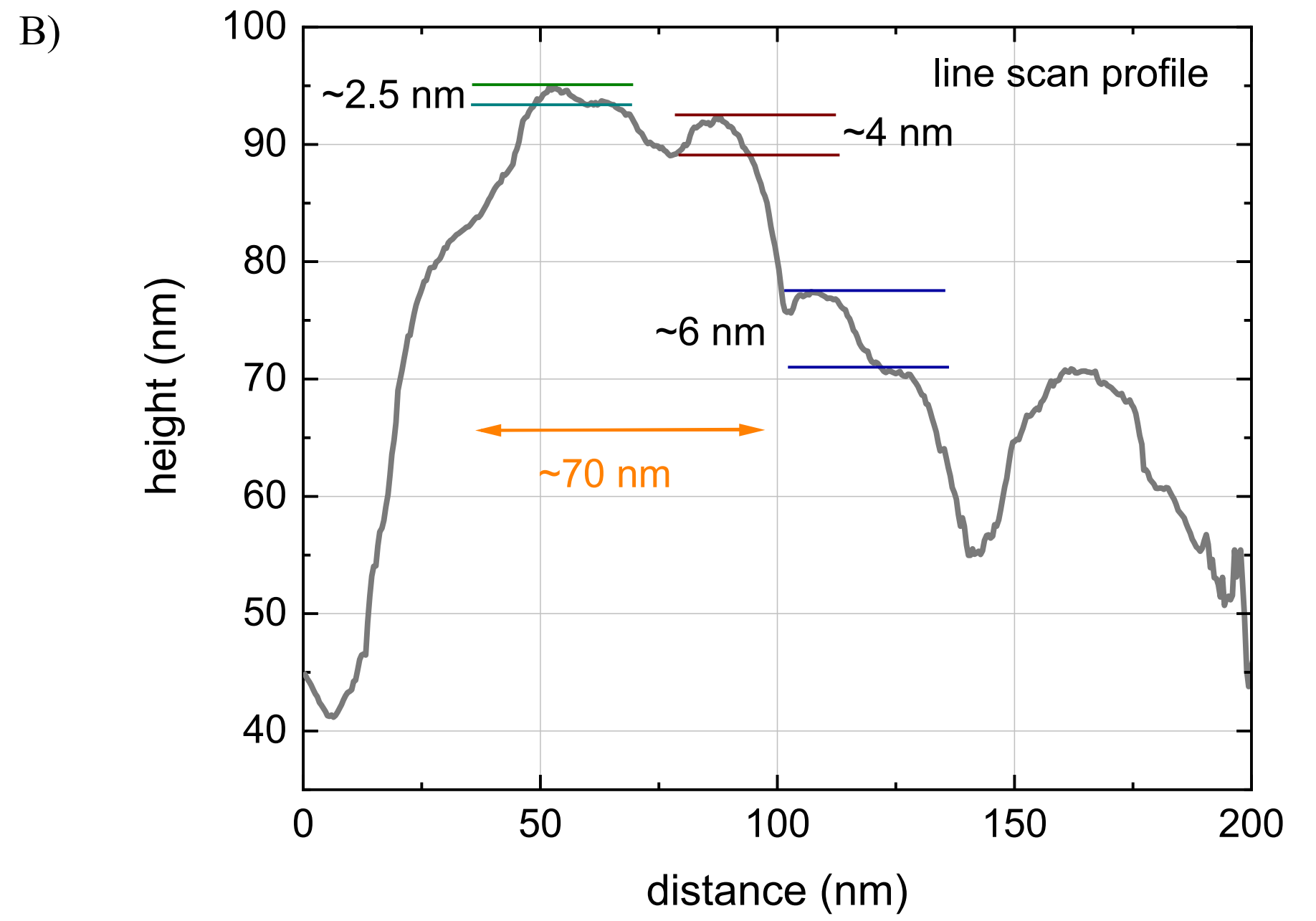

A) AFM scan image of the well-connected smaller germanium nanoparticles (~7nm) constituting the larger cauliflower spherical structures of about 50-100 nm. The AFM tip convolution should be considered in interpreting these graphs, as it may lead to an over estimation of the size in the lateral direction as well as to an underestimation of the full height of the particles. B) The sizes of the smaller nanoparticles can be inferred by profile scans (see the white line in A) measuring the height differences among the better separated particles.

FIG. 3

A)

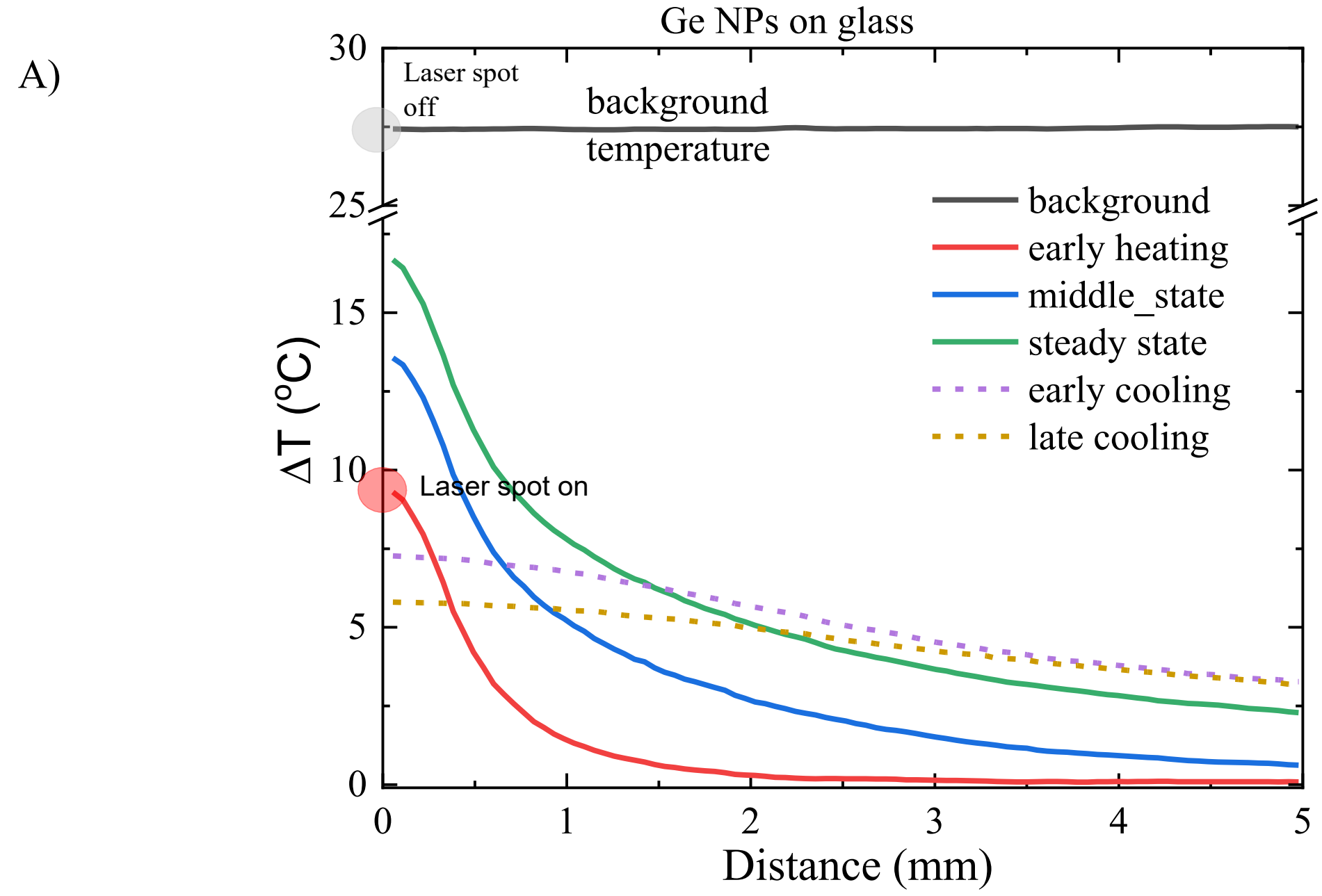

Ge NPs on glass
Laser spot off
background temperature
background
early heating
middle_state
steady state
early cooling
late cooling
Laser spot on
ΔT (°C)
Distance (mm)
30
25
15
10
5
0
0
1
2
3
4
5

B)

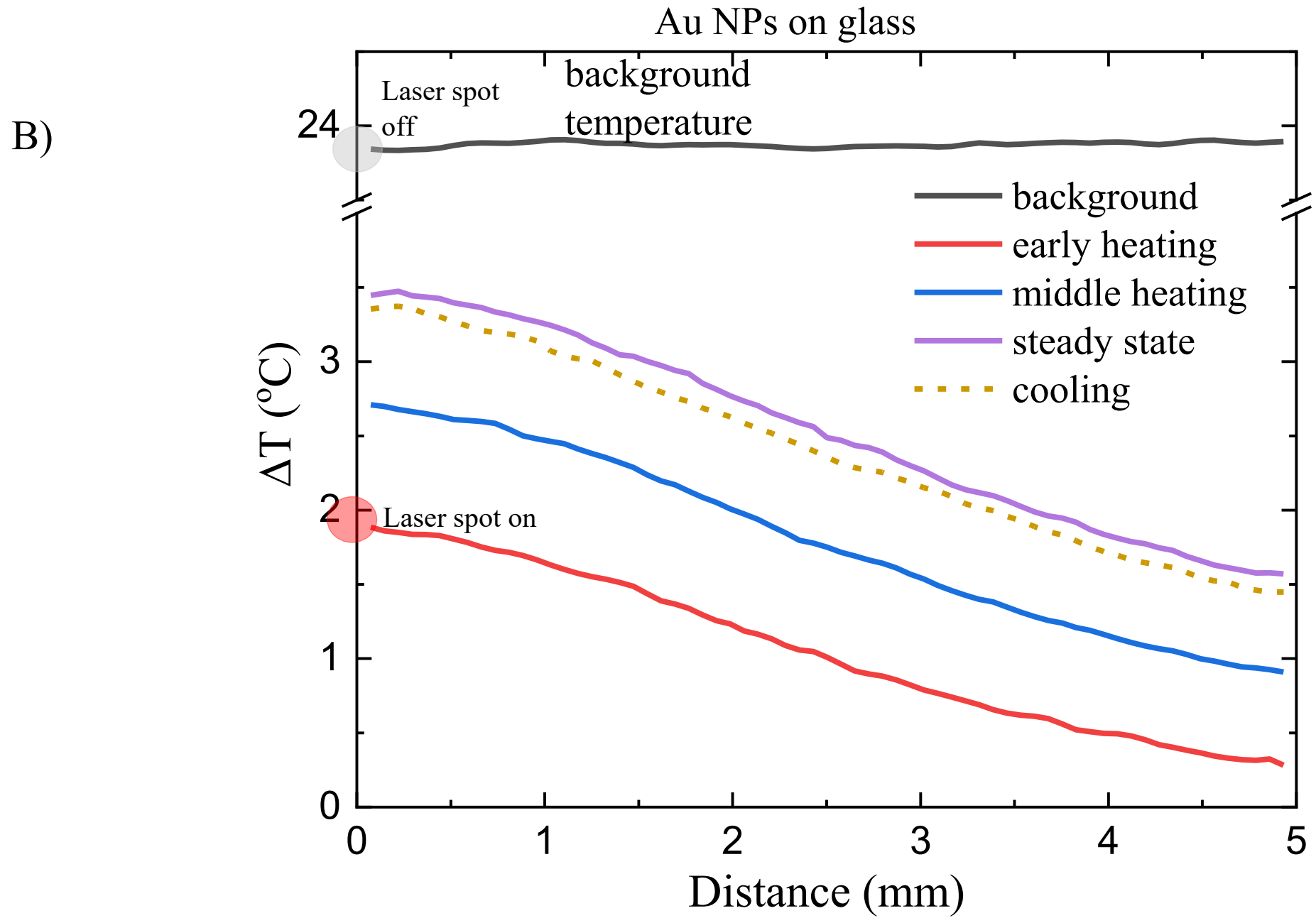

Au NPs on glass
Laser spot off
background temperature
background
early heating
middle heating
steady state
cooling
Laser spot on
ΔT (°C)
Distance (mm)
24
3
2
1
0
0
1
2
3
4
5

C)

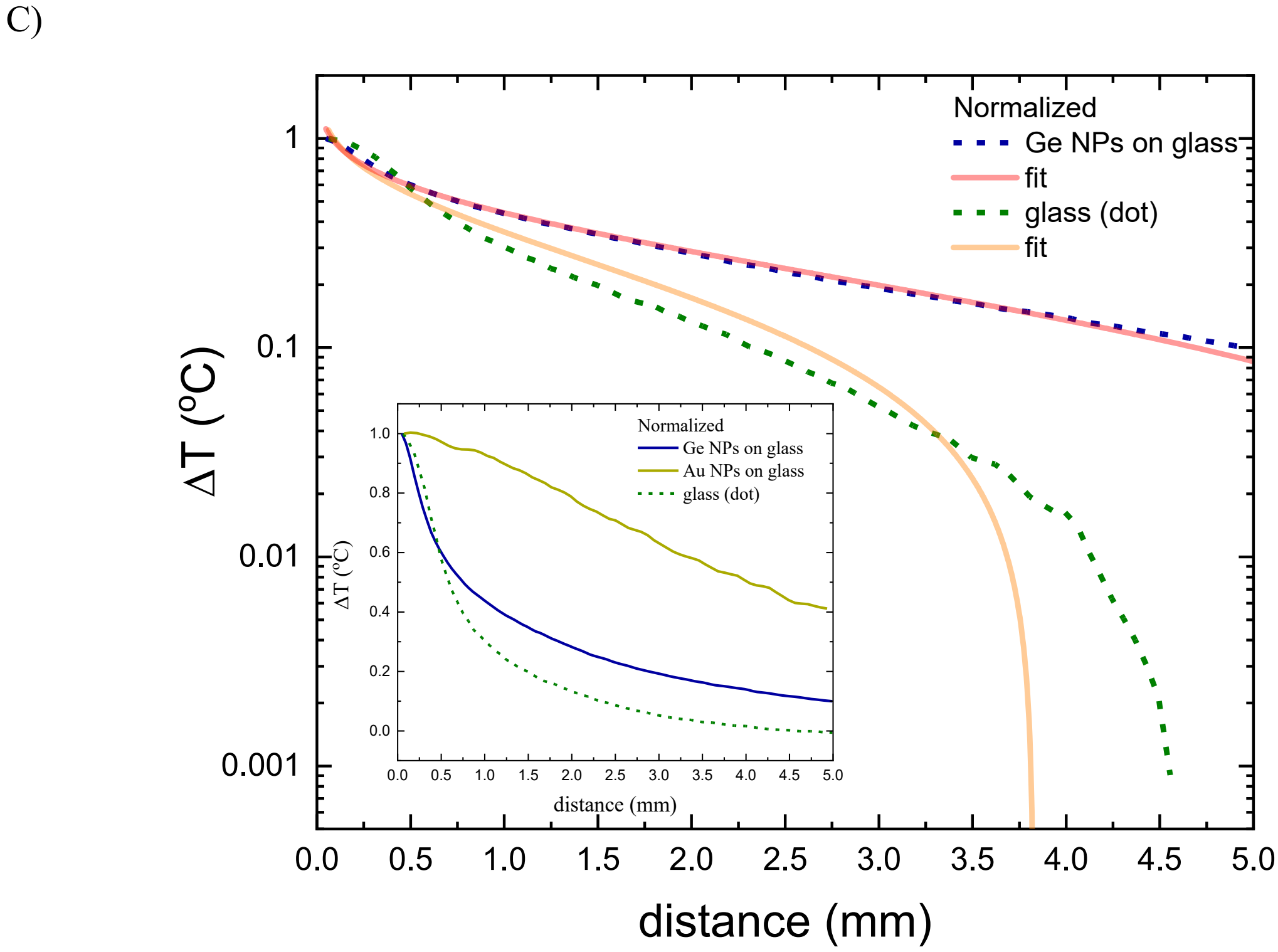

ΔT curves in the heating, steady state and cooling phase for (a) Germanium and (b) Gold nanoparticle assembled layers. Note that panels show the absolute value of the ambient temperature. A milder gradient in ΔT for gold nanoparticles is caused by their much higher thermal conductivity. (c) Normalized ΔT(x) for the Ge NPs on glass (blue dashed) and for glass substrate only with small ink spot to capture leaser heat (green dashed). The lateral heat flow in the germanium nanoparticles is significantly faster than in the glass substrate as can be seen from the smaller gradient and relatively higher overall steady state temperature. The curves were well fitted with Eq S25. The slower heat transport in the glass substrate alone allows for the distinct ΔT drop at larger distance (here 3.5 mm). The inset shows the same graphs with linear vertical scale.

FIG. 4

A)

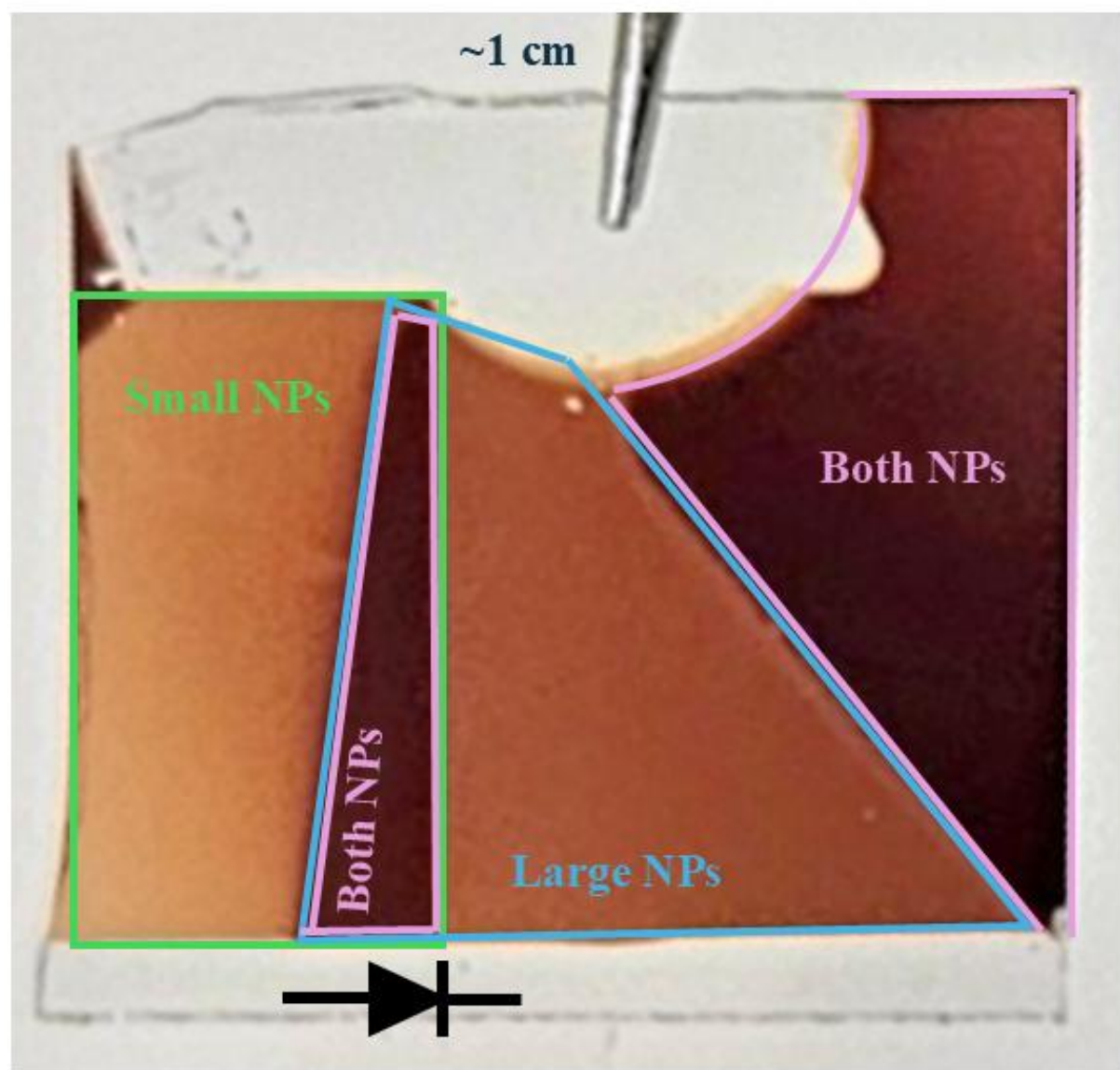

B)

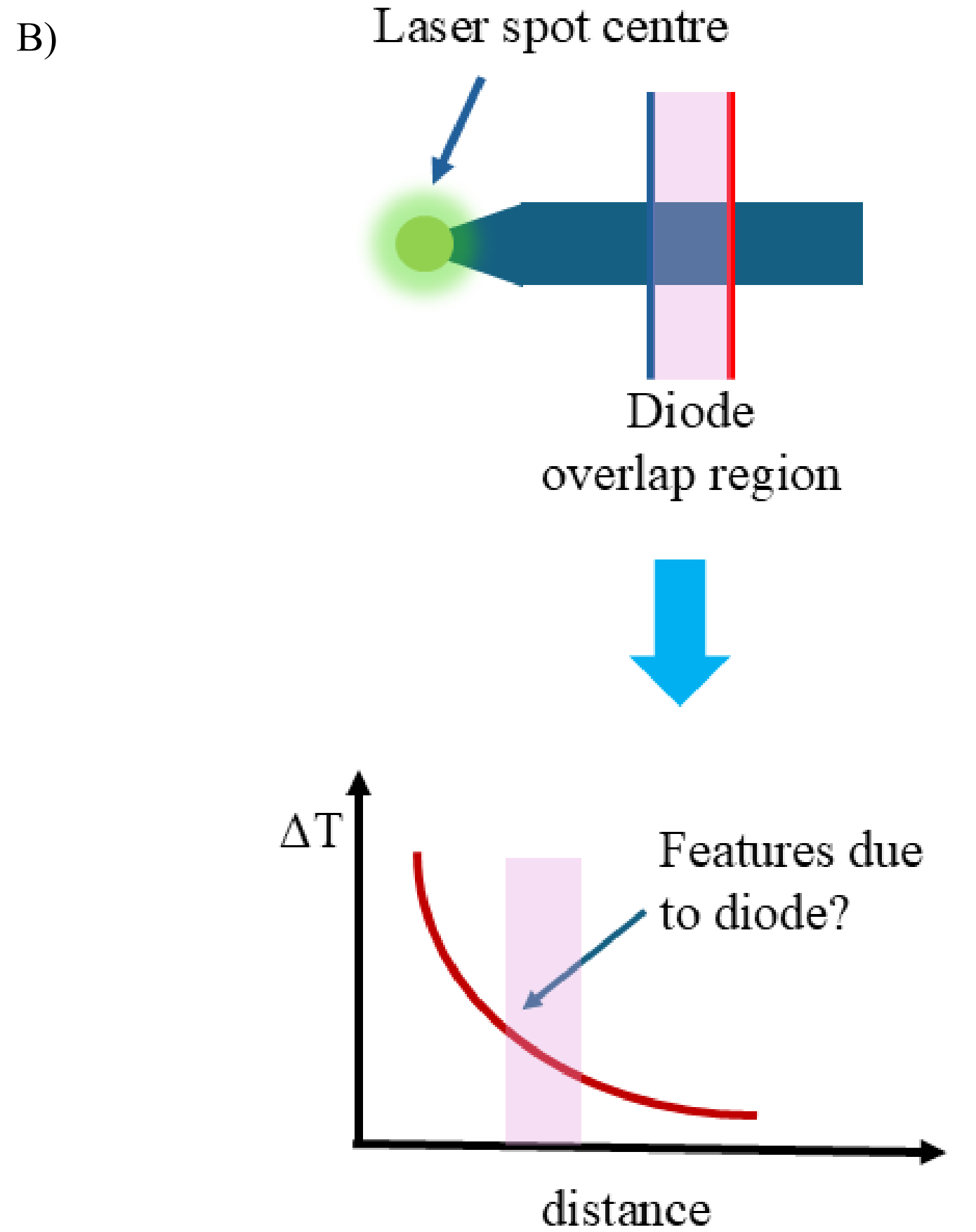

A) The Germanium nanoparticle sample, indicating regions composed of small (green) and large (blue) particles and their overlap area (pink). B) pencil mask to analyse the temperature versus distance from the laser spot heat source, resulting in the thermal profile graph.

FIG. 5

A)

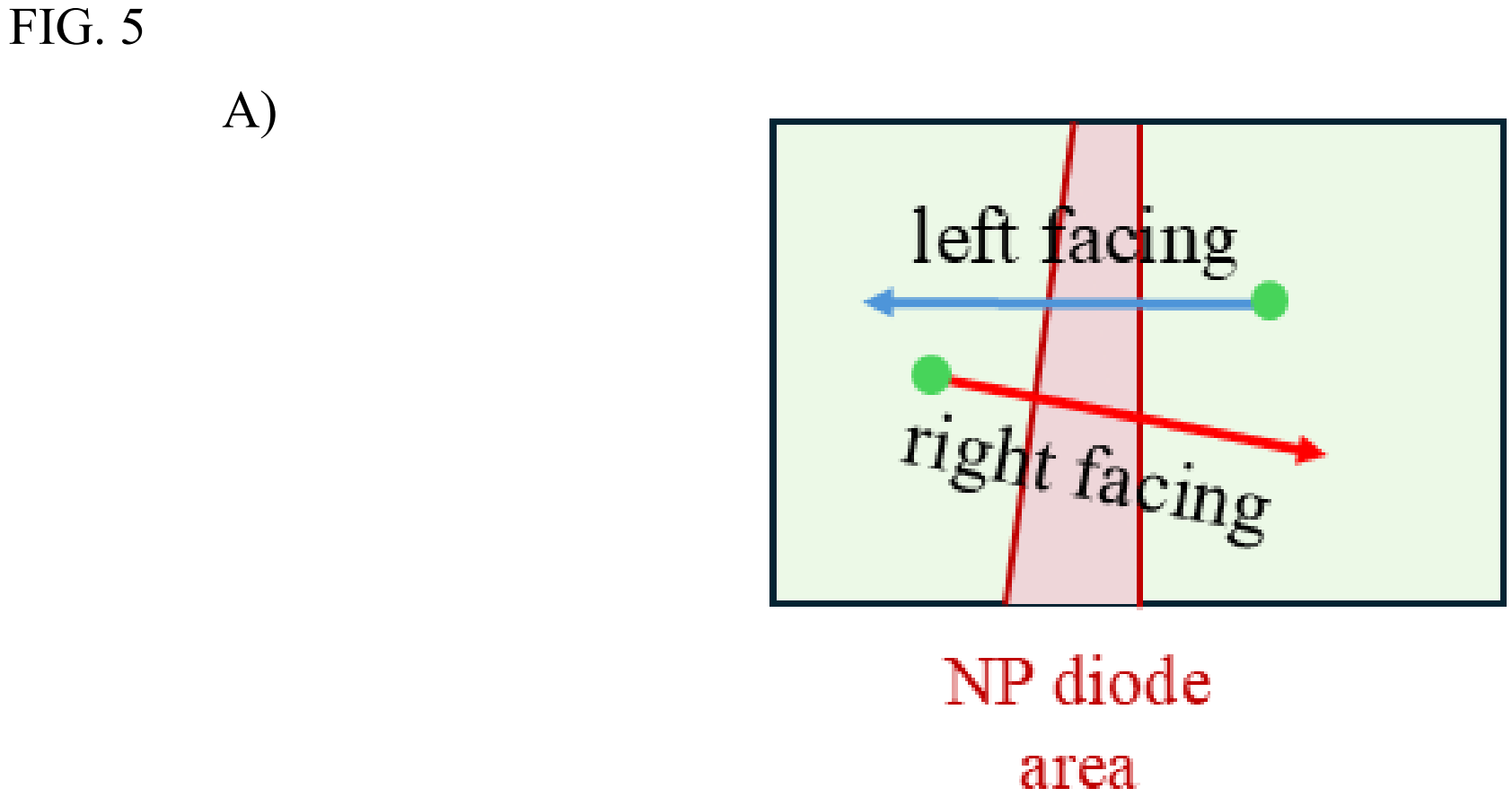

B)

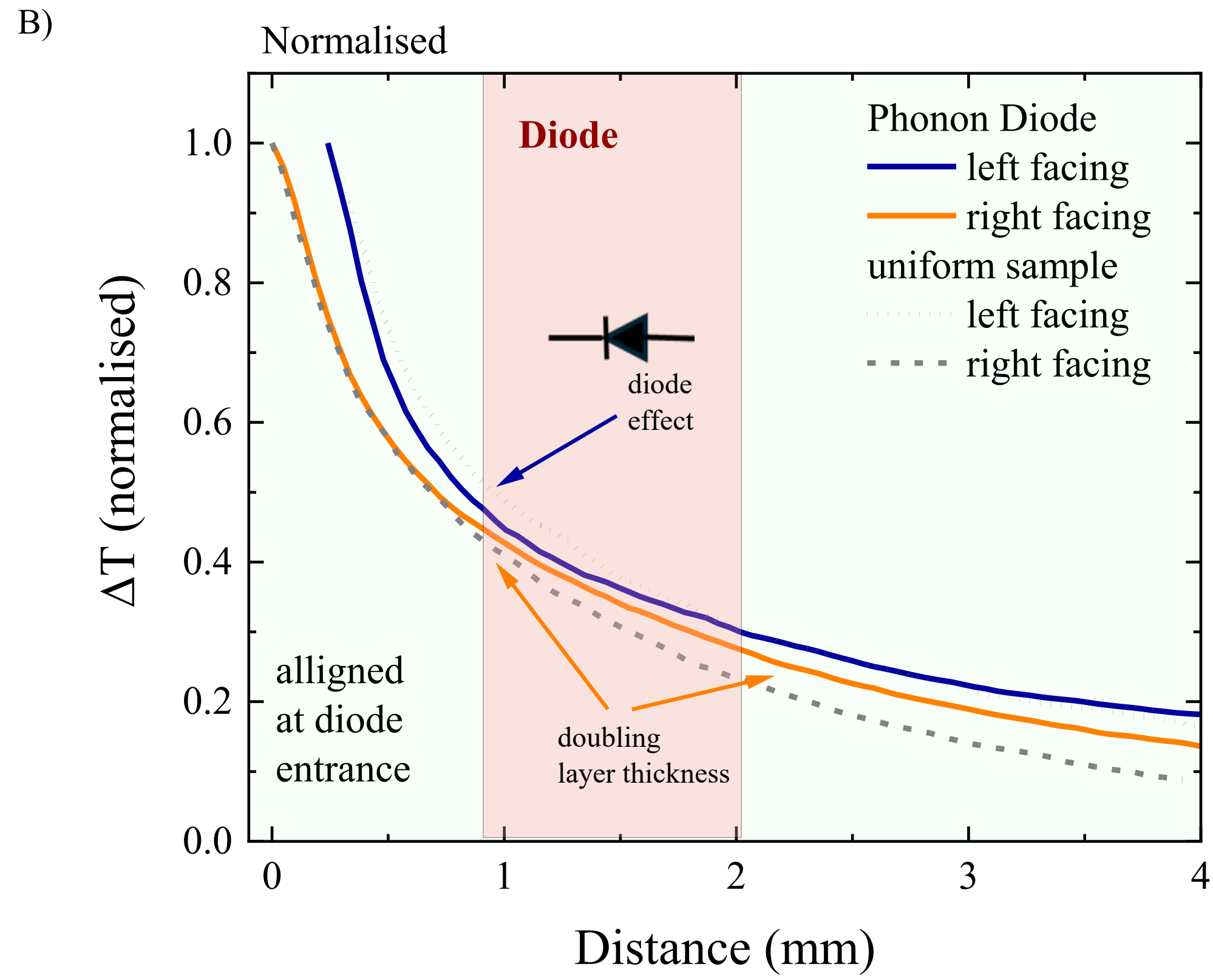

A) schematic depiction of germanium nanoparticle sample with diode region indicated in light red and heat flow analysis directions. The laser is incident on the red dots. B) ΔT as function of distance from laser spot for the germanium nanoparticles for the left facing (larger to smaller nanoparticles, blue) and right facing (smaller to larger nanoparticles, blue) heat flow. In the left facing case, ΔT is slightly lower than for a uniform layer, evidencing a phonon rectifying behaviour. For the right facing case, ΔT is slightly larger than for a uniform layer due to the double thickness of the germanium nanoparticle layer. The estimated rectifying strength is ~6%.

C)

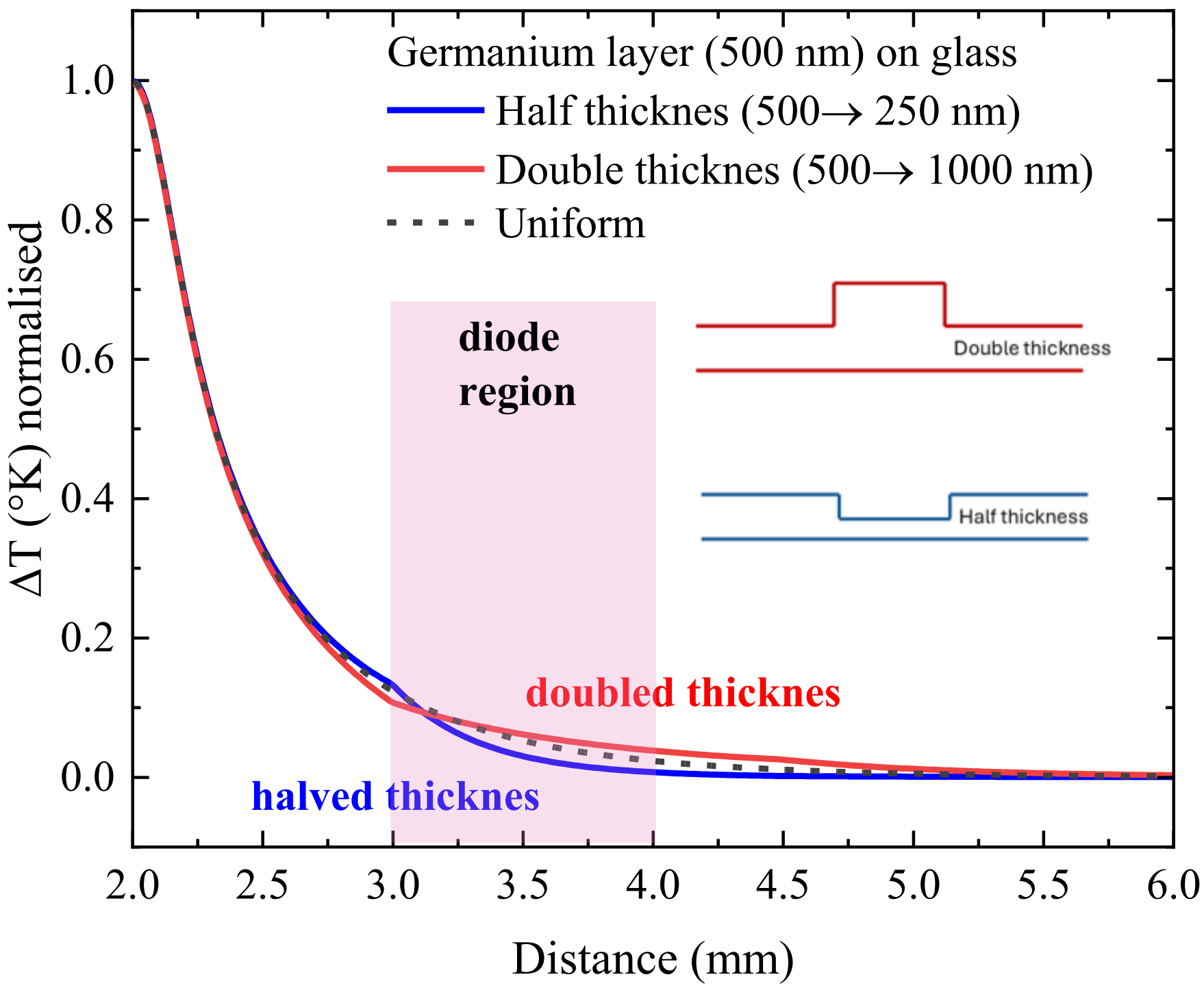

C) ΔT as function of distance from laser spot for the 500 nm thick Germanium layer on a glass substrate as calculated by solving the Fourier's law combined with energy conservation in 2D using an interface heat transport $G < 10$ W/m$^2$K. The overlap diode area between 3 and 4 mm distance from the Gaussian heat source, corresponds to the red curve for un-interrupted heat flow, which increases at the diode area (pink area) due to the doubling of the thickness, leading to higher temperatures. The phonon diode is mimicked by halving the layer (blue) in thickness at the diode area (pink), leading to reduced hat flow and lower temperatures. Due to the larger heat flow in the double layer (red), the temperatures just before the increased or reduced thickness are slightly lower or higher, due to stronger flowing or accumulated heat. The black dashed line corresponds to a uniform germanium layer. The spatial extend of the mimicked diode is shown in the inset.